\documentstyle[12pt]{article}
\global\parskip 6pt
\begin{document}
%%%%%HERE ARE SOME GENERAL DEFINITIONS AND COMMANDS
\newcommand{\nc}{\newcommand}
\newcommand{\rnc}{\renewcommand}
%\catcode`\@=11                                         %this numbers
%\@addtoreset{equation}{section}                        %equations
%\rnc{\theequation}{\arabic{section}.\arabic{equation}} %by sections
\nc{\be}{\begin{equation}}
\nc{\ee}{\end{equation}}
\nc{\bea}{\begin{eqnarray}}
\nc{\eea}{\end{eqnarray}}
%%%%%%GREEK LETTERS AND SUCH
\rnc{\a}{\alpha}
\nc{\g}{\gamma}
\rnc{\d}{\delta}
\nc{\e}{\eta}
\nc{\eb}{\bar{\eta}}
\nc{\f}{\phi}
\nc{\fb}{\bar{\phi}}
\nc{\p}{\psi}
\rnc{\pb}{\bar{\psi}}
\rnc{\c}{\chi}
\nc{\cb}{\bar{\c}}
\nc{\m}{\mu}
\nc{\n}{\nu}
\rnc{\o}{\omega}
\rnc{\t}{\theta}
\nc{\tb}{\bar{\theta}}
\nc{\M}{{\cal M}}                           %SOME MODULI SPACE
\nc{\C}{{\cal A}/{\cal G}}                  %SPACE OF GAUGE ORBITS
\nc{\A}[1]{{\cal A}^{#1}/{\cal G}^{#1}}     %SPACE OF GAUGE ORBITS
\nc{\RC}{{\cal R}_{\C}}                     %CURVATURES
\nc{\RM}{{\cal R}_{\M}}                     %----------
\nc{\RX}{{\cal R}_{X}}
\nc{\RY}{{\cal R}_{Y}}
\nc{\ra}{\rightarrow}
\nc{\ot}{\otimes}
\rnc{\ss}{\subset}
\rnc{\lg}{{\bf g}}                          %THE LIE ALGEBRA OF G
\nc{\cs}{\c_{s}}                            %REGULARIZED EULER NUMBER
\nc{\del}{\partial}
\nc{\dx}{\dot{x}}
\rnc{\O}[2]{\Omega^{#1}({#2},\lg)}          %LIE ALGEBRA VALUED FORMS
\def\tft{top\-o\-log\-i\-cal field the\-o\-ry}
\def\tgt{top\-o\-log\-i\-cal gauge the\-o\-ry}
\def\sqm{su\-per\-sym\-met\-ric quant\-um mech\-an\-ics}
%%%%%%%%%%%%%%%%%%%%%%%%%%%%%%%%%%%%%%%%%%%%%%%%%%%%%%%%%%%%%
\begin{titlepage}
\newlength{\titlehead}
\settowidth{\titlehead}{NIKHEF-H/91-???}
\begin{flushright}
\parbox{\titlehead}{
\begin{flushleft}
MZ-TH/91-42\\
December 1991\\
\end{flushleft}
}
\end{flushright}
\begin{center}
{\Large\bf Topological Gauge Theories}\\
\vskip .1in
{\Large\bf from Supersymmetric Quantum Mechanics}\\
\vskip .1in
{\Large\bf on Spaces of Connections}\\
\vskip .3in
{\bf Matthias Blau}\footnote{e-mail: t75@nikhefh.nikhef.nl, 22747::t75} \\
\vskip .10in
{\em NIKHEF-H}\\
{\em P.O. Box 41882, 1009 DB Amsterdam}\\
{\em The Netherlands}
\vskip .15in
{\bf George Thompson}\footnote{e-mail: thompson@vipmzt.physik.uni-mainz.de}\\
\vskip .10in
{\em Institut f\"ur Physik} \\
{\em Johannes-Gutenberg-Universit\"at Mainz}\\
{\em Staudinger Weg 7, D-6500 Mainz, FRG}\\
\end{center}
\begin{abstract}
We rederive the recently introduced $N=2$ topological gauge theories,
representing the Euler characteristic of moduli spaces ${\cal M}$ of
connections, from supersymmetric quantum mechanics on the infinite
dimensional spaces ${\cal A}/{\cal G}$ of gauge orbits. To that end
we discuss variants of ordinary supersymmetric quantum mechanics which have
meaningful extensions to infinite-dimensional target spaces and introduce
supersymmetric quantum mechanics actions modelling the Riemannian geometry of
submersions and embeddings, relevant to the projections ${\cal A}\rightarrow
{\cal A}/{\cal G}$ and inclusions ${\cal M}\subset{\cal A}/{\cal G}$
respectively. We explain the relation between Donaldson theory and the
gauge theory of flat connections in $3d$ and illustrate the general
construction by other $2d$ and $4d$ examples.
\end{abstract}
\end{titlepage}

%\tableofcontents
\setcounter{footnote}{0}

\section{Introduction}

{}From its historical development it is evident that \tft\
is closely related to \sqm\ with infinite dimensional target
spaces. Nevertheless this aspect of \tft\ has attracted little attention
in subsequent developments and the
purpose of this paper is to fill this gap and to illustrate the usefulness
of this perspective.

In particular, we will use \sqm\ on the space $\C$ of gauge orbits of
connections to give a rather pedestrian derivation of the
\tgt\ models introduced recently in \cite{btmq} and discussed there from
several other points of view (e.g.~via the Mathai-Quillen formalism
\cite{mq,aj} and in terms of topological $N=2$ superfields). The present
derivation is aimed at clarifying the origin of the fundamental property
of these theories (obviously reminiscent of \sqm): that the partition
function of the action $S_{\M}$
associated to a given moduli space $\M$ of connections equals the Euler
characteristic $\c(\M)$ of $\M$,
\be
Z(S_{\M})=\c(\M)\label{1}\;\;.
\ee
Moreover, as a side-result, we will also see in this paper that
the action of Donaldson theory \cite{ewdon}, the prototype of a
cohomological \tft\ (see \cite{pr} for a review), can be identified
term by term with the standard action $S_{X,V}$ of \sqm\ on a manifold
$X$ coupled
to a potential $V$ for $X=\A{3}$ (the space of gauge orbits in three
dimensions) and $V=CS$ (the Chern-Simons functional). In view of the
considerations in \cite{at} this is almost tautologically true, but we
have included this result here because we have never seen it spelled out
explicitly (i.e.~in terms of the Riemannian metric, connection, and
curvature of $\A{3}$)
and a number of things can be learned from this construction.

The main reason why this way of constructing and looking at topological
gauge theories has enjoyed only limited popularity at best,
is that Donaldson theory has some features which are not available in general.
In particular, the existence of a potential on $\A{3}$, the Chern-Simons
functional, which leads to a description of moduli spaces
$\M_{I}\subset\A{4}$ of instantons in $4d$ (via its gradient flow) is a
fortuitous coincidence which reflects the richness of Donaldson theory.
The fact that, used in this way, supersymmetric quantum mechanics on $\A{n}$
leads to a theory in $n+1$ dimensions, in general prevents one from applying
this method to the construction of topological gauge theories based on
moduli spaces $\M\subset\A{n+1}$ as no suitable potential function on
$\A{n}$, nor any other obvious means of exerting control over $\M$ from an
$n$-dimensional point of view, will exist.

In this letter we show that the $n$-dimensional topological gauge theory
actions $S_{\M}$ associated with a given moduli space $\M\subset\A{n}$ can
be constructed directly from \sqm\ on $\A{n}$, the intermediate theory on
$\A{n+1}$ playing only an auxiliary role. Not only is this method manifestly
covariant (in contrast with the above construction which leads to a
$(3+1)$-dimensional description of Donaldson theory), but also (and more
importantly) it frees us from the necessity of having to find an
$n$-dimensional description of a moduli space of connections in $n+1$
dimensions. This also completes the circle of ideas relating Floer cohomology,
the Chern-Simons functional, instanton moduli spaces, the Casson invariant,
and moduli spaces of flat connections from the point of view of \cite{btmq}.

Our construction will be based on the standard localization and index theory
arguments of \sqm. In particular, we will make use of the fact that the
theory $S_{X,V}$ localizes to the critical point set $X_{V}$ of $V$ to derive
a topological gauge theory of flat connections in three dimensions (the
critical points of $CS$) from Donaldson theory. In order to describe moduli
spaces of connections which are not of the form $X_{V}$ for some $V$
we introduce a new variant of \sqm\ in finite dimensions which localizes
onto an arbitrary given submanifold $Y\ss X$ and describes the Riemannian
geometry of embeddings via the Gauss-Codazzi equations. The corresponding
action $S_{Y\ss X}$ also makes sense when applied to $X=\A{n}$ and any
finite-dimensional moduli subspace $Y=\M$ of $\A{n}$.

In section 2 we review the pertinent features of \sqm\ (path integral
representation of the Euler characteristic, evaluation of the partition
function of $S_{X,V}$, localization) and introduce the Gauss-Codazzi
quantum mechanics action $S_{Y\ss X}$. In section 3 we describe the
Riemannian geometry of (moduli) spaces of connections and its field-theoretic
realization. We also establish the relation between Donaldson theory and
\sqm\ on $\A{3}$, and construct the topological gauge theory of flat
connections in three dimensions. In section 4 we illustrate the use of
$S_{Y\ss X}$ by treating moduli spaces of instantons and of flat
connections in two dimensions.

\section{Review of supersymmetric quantum mechanics}

The fundamental action of \sqm\, modelling the de Rham complex of a
Riemannian manifold $(X,g)$, is (we are using the conventions of \cite{pr})
\be
S_{X} =  \int_{0}^{\beta} dt [ i \dx^{\m}B_{\m} +
\frac{1}{2}g^{\m\n}B_{\m}B_{\n}+\frac{1}{4}
R^{\m\n}_{\;\;\rho\sigma}\bar{\psi}_{\m}\psi^{\rho}\bar{\psi}_{\n}
\psi^{\sigma}  -i \bar{\psi}_{\m}\nabla_{t}\psi^{\m} ] \;\;.
\label{2}
\ee
Here $x^{\m}$ are the coordinates of the Riemannian manifold $(X,g)$
with curvature tensor $\RX= (R^{\m\n}_{\;\;\rho\sigma})$,
$\psi^{\m}$ and $\bar{\psi}_{\m}$ are Grassmann odd
coordinates, and
the covariant derivative $\nabla_{t}$ is the pull-back of the covariant
derivative on $X$ to the one dimensional space with Euclidean time
coordinate $t$.
Upon integrating out the auxiliary field $B$, one recovers the action of
\cite{ewqm,ewqm1,ag1,ag2} with the spinors appearing there
decomposed into their components. We choose $\bar{\psi}$ and
$\psi$ to be independent real fields (instead of complex conjugates).
The supersymmetry of the action (\ref{2}) is
\bea
\d x^{\m} &=& \p^{\m}\;,\;\;\;
\d\pb_{\m}=B_{\m}-\Gamma^{\n}_{\;\m\rho}\pb_{\n}\p^{\rho}\nonumber\\
\d \p^{\m}&=& 0\;\;,\;\;\;\;
\d B_{\m}=\Gamma^{\n}_{\;\m\rho}B_{\n}\p^{\rho}
            -\frac{1}{2}R^{\n}_{\;\m\rho\sigma}\pb_{\n}\p^{\rho}\p^{\sigma}
\label{susy}\;\;.
\eea
$S_{X}$ can be written as the supersymmetry variation of
$\int_{0}^{\beta}dt[\pb_{\m}(i\dx^{\m}+\frac{1}{2}g^{\m\n}B_{\n})]$
and this has far-reaching consequences. In particular, reinterpreting
$\d$ as a BRST operator, this demonstrates that the ground state
reduction of \sqm\ is topological and that the theory is independent of the
coefficient of $B^{2}$ (cf.~\cite[pp.~140-176]{pr} for
a detailed discussion of \sqm\ in the context of topological field
theories).

As is well known, the partition function $Z(S_{X})$ of (\ref{2})
with periodic boundary conditions on {\em all} the fields is the
Euler number $\c(X)$ of $X$. The way to see this is to start with the
definition of $\c(X)$ as the Euler characteristic of the de Rham complex
of $X$, $\c(X)=\sum_{k}(-1)^{k}b_{k}(X)$ (where
$b_{k}(X)=\dim H^{k}(X,{\bf R})$ is the $k$'th Betti number of $X$)
and to rewrite this as the
Witten index $\c(X)=tr(-1)^{F}\exp(-\beta H)$
of the Laplace operator $H\equiv\Delta=dd^{*}+d^{*}d$ on differential forms.
One then uses the
Feynman-Kac formula to represent this as a supersymmetric path integral
with the action (\ref{2}) and periodic boundary conditions on the
anticommuting variables $\p^{\m}$ (due to the insertion of $(-1)^{F}$).

The partition function $Z(S_{X})$ can be evaluated explicitly to give
a path integral proof of the Gauss-Bonnet theorem which expresses $\c(X)$
as an integral over $X$ of the Pfaffian of the curvature $\RX$,
\be
Z(S_{X})=\c(X)=\int_{X}P\!f(\RX)\label{5}\;\;.
\ee
The crucial
fact responsible for the reduction of the integral over the loop space $LX$
of $X$ (the path integral) to an integral over $X$ (the Gauss-Bonnet integral)
is the $\beta$-independence of the Witten index. This permits one to evaluate
the partition function in the limit where the radius $\beta$ of the circle
tends to zero. In this limit it can be seen that only the Fourier zero modes
(e.g.~$\dx=0$) of the fields are relevant, the contributions from the
other modes cancelling identically between the bosonic and fermionic fields.
All this is, of course, also an immediate consequence of the BRST symmetry
and topological nature of \sqm\ mentioned above.
It is the analogue in infinite dimensions of this observation that allows us
to construct topological gauge theories in $n$ (instead of $n+1$) dimensions
from \sqm\ on $\A{n}$.

As they stand, the partition function of (\ref{2}) and the right hand side of
(\ref{5}) do not make sense for infinite dimensional target spaces.
There are, however, two generalizations of (\ref{2})
which turn out to have meaningful counterparts on $\C$. The first of these
involves a choice of potential $V(x)$ on $X$. The corresponding action
\bea
S_{X,V}& = & \int dt [ i(\dx^{\m}+sg^{\m\n}\del_{\n}V(x))B_{\m} +
\frac{1}{2}g^{\m\n}B_{\m}B_{\n}+\frac{1}{4}
R^{\m\n}_{\;\;\rho\sigma}\bar{\psi}_{\m}\psi^{\rho}\bar{\psi}_{\n}\psi^{\sigma}
\nonumber\\
& & -i \bar{\psi}_{\m}(\d^{\m}_{\n}\nabla_{t}+sg^{\m\rho}
\nabla_{\rho}\del_{\n}V)\psi^{\n} ] \;\;.
\label{6}
\eea
($s$ is a parameter) arises by replacing the exterior derivative $d$ by
$d_{sV}\equiv\exp(-sV)d\exp(sV)$. As there is a one-to-one correspondence
between $d$- and $d_{sV}$-harmonic forms this also represents $\c(X)$
(independently of $s$). In this case the additional freedom in the
choice of $s$ allows one to reduce $Z(S_{X,V})$ to an integral over the
set of critical points of $V$ in the limit $s\ra\infty$ (alternatively
one uses the fact that the partition function is also independent of the
coefficient of $B^{2}$ and observes that $\dx^{\m}+sg^{\m\n}\del_{\n}V(x)=0$
implies $\dx^{\m}=\del_{\n}V(x)=0$ for any $s\neq 0$ by squaring and
integrating). In the case that the critical points of $V$ are isolated
and non-degenerate one arrives at the classical Poincar\'e-Hopf-Morse
theorem
\be
\c(X)=\sum_{x_{k}: dV(x_{k})=0}(\pm 1)\label{7}
\ee
which calculates $\c(X)$ as the signed sum of critical points of $V$.
More generally (\ref{7}) holds, and can be derived from \sqm\, for
the sum over the zeros of a generic vectorfield on $X$.
If the critical points are not isolated then, by a combination of the
arguments leading to (\ref{5}) and (\ref{7}), one finds
\be
\c(X)=\sum_{(k)}\c(X_{V}^{(k)})\label{8}\;\;,
\ee
where the $X_{V}^{(k)}$ are the connected components of the critical point
set of $V$. The relevance of this for our purposes is that the right hand side
of (\ref{8}) may be well defined, even if $X$ is infinite dimensional,
provided that $X_{V}$ is finite dimensional. In that case $\c(X_{V})$ is
well defined and can be regarded as a regularized Euler number of $X$
(this is the point of view adopted in \cite{aj}). The advantage of our
construction is that it permits an {\em a priori} identification of
this $V$-dependent regularized Euler number of $X$ with the Euler number
of $X_{V}$.

Although this looks like a satisfactory state of affairs, we may not always
be so fortunate to have a potential at our disposal whose critical points
define precisely the (moduli) subspace $Y\subset X$ we are interested in.
In fact, it follows from (\ref{8}) that in finite dimensions
$\c(Y)=\c(X)$ is a necessary condition for this to be possible. Moreover,
it is by now well known that even on a compact four-manifold there are
critical points of the vacuum Yang-Mills functional other than instantons.
Thus a suitable potential is unlikely to exist e.g.~for the
instanton moduli spaces $\M_{I}\subset\A{4}$. We thus require a generalization
of (\ref{2}) which calculates the Euler number $\c(Y)$ for any submanifold
$Y\subset X$ regardless of whether $Y=X_{V}$ for some $V$ or not.

In that setting we have the classical Gauss-Codazzi equations which
relate the intrinsic curvature $\RY$ of $Y$ (with the induced metric)
to $\RX$ restricted to $Y$ and the extrinsic curvature
(second fundamental form) of the embedding $i:Y\hookrightarrow X$.
The second fundamental form $K_{Y}$ of $(Y,i)$ is defined by
$K_{Y}(v,w)=(\nabla_{i_{*}v}i_{*}w)^{\perp}$,
where $v,w\in TY$, $\nabla$ is the Levi-Civit\`a connection on $X$, and
$(.)^{\perp}$ denotes projection onto the normal bundle
to $TY$ in $TX|_{Y}$. If $Y$ is a hypersurface in $X$, this
reduces to the more mundane statement that the extrinsic curvature
is essentially the normal derivative of the induced metric.
The Gauss equation now states that
\bea
\langle{\cal R}_{Y}(u,v)z,w\rangle &=& \langle{\cal R}_{X}(u,v)z,w\rangle
\nonumber\\&+&
(\langle K_{Y}(v,z),K_{Y}(u,w)\rangle -(u\leftrightarrow v))
\label{10}\;\;.
\eea
Our construction of an action $S_{Y\ss X}$ calculating $\c(Y)$ via
the Gauss-Bonnet theorem applied to (\ref{10}) will be modelled on
(\ref{10}) itself. Essentially, it will consist of the action $S_{X}$
(\ref{2}) plus a Lagrange multiplier term enforcing the restriction to
$Y\ss X$. Provided that this restriction is performed in a way consistent
with the supersymmetries of de Rham \sqm\
this will automatically give rise to the second term of (\ref{10}).

More concretely, assume that $Y\ss X$ is (locally) given by
\[Y=\{x\in X: F^{a}(x)=0,\;\;a=1,\ldots,dim(X)-dim(Y)\}\]
(the relation between the formulae arising from this implicit description
and that in terms of an explicit embedding $y^{k}(x^{\m})$ is explained
e.g.~in \cite{bc}). We then group the fields appearing in (\ref{2}) into
a topological $N=2$ superfield
\be
X^{\m}(t,\theta,\bar{\theta})
=x^{\m}(t)+\theta \p^{\m}(t) + \bar{\theta}
g^{\m\n}\pb_{\n}(t) - \theta \bar{\theta}(g^{\m\n} B_{\n}(t)
+g^{\beta\n}\Gamma^{\m}_{\;\beta\lambda}\pb_{\n}\p^{\lambda})
\label{11}
\ee
($\t$ and $\tb$ are Grassmann odd scalars). This choice of superfields is
designed to reproduce the supersymmetry transformations (\ref{susy}). As we
will see below, the second term of the $\t\tb$-component moreover
leads to a Taylor expansion of superfields in terms of covariant derivatives
so that superspace actions are manifestly covariant. We also introduce
$N=2$ Lagrange multiplier fields
\be
\Lambda_{a}(t,\theta,\bar{\theta}) =  \lambda_{a}(t) +  \theta \sigma_{a}(t) +
\bar{\theta} \bar{\sigma}_{a}(t)  +  \theta \bar{\theta} b_{a}(t) \;\;,
\label{12}
\ee
and choose the action to be
\be
S_{Y\ss X}=S_{X}+\a\int dt\int d\t\,d\tb
\Lambda_{a}(t,\t,\tb)F^{a}(X(t,\t,\tb))\label{13}\;\;,
\ee
so that the integration over the $\Lambda_{a}$ imposes the superconstraints
$F^{a}(X)=0$. The argument given above leading to the elimination of the
non-constant modes is not affected by the addition of this term and thus,
upon Taylor expanding $F^{a}(X)$, (\ref{13}) becomes (all `fields' are now
time independent)
\bea
S_{Y\ss X} &=&\beta[\frac{1}{2}g^{\m\n}B_{\m}B_{\n}+\frac{1}{4}
R^{\m\n}_{\;\;\rho\sigma}
\bar{\psi}_{\m}\psi^{\rho}\bar{\psi}_{\n}\psi^{\sigma}]\label{14}\\
&+&\a[b_{a}F^{a}-\sigma_{a}\pb^{\m}\del_{\m}F^{a}+
\bar{\sigma}_{a}\p^{\m}\del_{\m}F^{a}+\lambda_{a}
(B^{\m}\del_{\m}F^{a}-g^{\beta\n}\nabla_{\beta}\del_{\lambda}F^{a}
\pb_{\n}\p^{\lambda})]\;\;.\nonumber
\eea
We see that the integral over $b$ restricts the bosonic coordinates to $Y$
while the integrals over $\sigma$ and $\bar{\sigma}$ constrain the fields
$\p^{\m}$ and $\pb_{\m}$ to be tangent to $Y$. It is now a simple matter
to perform the Gaussian integrals over the remaining auxiliary fields
$B_{\m}$ and $\lambda_{a}$ with the result
\be
S_{Y\ss X}=(\frac{1}{4}R^{\m\n}_{\;\;\rho\sigma}+\frac{1}{2}
g^{\m\a}g^{\n\beta}\nabla_{\a}\del_{\rho}F^{a}(F^{-1})_{ab}
\nabla_{\beta}\del_{\sigma}F^{b})\pb_{\m}\p^{\rho}\pb_{\n}\p^{\sigma}\;\;.
\label{15}
\ee
Here $F^{ab}$ is the matrix
$F^{ab}=\del_{\m}F^{a}\del_{\n}F^{b}g^{\m\n}$
and the description of $Y\ss X$ in terms of the $F^{a}$ is valid at points
where $\det(F^{ab})\neq 0$ so that $F^{ab}$ is indeed invertible there.
We see that $\a$ has dropped out (as it should) and we have rescaled the
$\pb$'s by $\beta^{1/2}$ to eliminate all $\beta$-dependence from both the
measure and the action. Equation (\ref{15}) is precisely the Gauss equation
(\ref{10}) which we have thus derived from \sqm. Therefore, upon expanding
the path integral to soak up the $dim(Y)$ fermionic $\p$ and $\pb$ zero modes,
we will indeed find
\be
Z(S_{Y\ss X})=\c(Y)\label{16}\;\;,
\ee
now valid for arbitrary submanifolds $Y\ss X$ (not necessarily of the form
$X_{V}$). This is the generalization we need to be able to apply
\sqm\ to spaces of connections.
We also see that, in a certain sense, the action $S_{X,V}$
(\ref{6}) is a special case of the action $S_{Y\ss X}$ (\ref{13}), the zero
mode of $B$ playing the role of the multiplier $b$.

Finally we mention that one can also construct \sqm\ actions $S_{Z\ra X}$
for Riemannian submersions $Z\ra X$ instead of embeddings, deriving the
O'Neill equations \cite{on} in this case instead of the Gauss-Codazzi
equations. This is most effortlessly done when the submersion is actually
a fibration. Instead of developing the full mashinery here, we will
illustrate this in passing in the following section.

\section{Donaldson theory and flat connections in $3d$}

We will now introduce the data entering into the construction of
the \sqm\ actions $S_{X,V}$ and $S_{Y\ss X}$ on spaces of connections
(see e.g.~\cite{nr,ims,bv}). Let $(N,g)$ be a compact, oriented, Riemannian
$n$-manifold, $P\ra N$ a principal $G$ bundle over $N$, $G$ a compact
semisimple Lie group and $\lg$ its Lie algebra. We denote by $\cal A$ the
space of (irreducible) connections on $P$,
by $\cal G$ the infinite dimensional gauge group of vertical automorphisms
of $P$ (modulo the center of $G$), by $\O{k}{N}$ the space of $k$-forms on
$N$ with values in the adjoint bundle $ad\,P:=P\times_{ad}\lg$ and by $d_{A}$
the covariant exterior derivative. The spaces $\O{k}{N}$
have natural scalar products defined by the metric $g$ on $N$ (and the
corresponding Hodge operator $*$) and an invariant scalar product $tr$ on
$\lg$, namely
\be
\langle X,Y \rangle = \int_{M}tr(X*Y)\;\;,\;\;\;\;\;\;\;\;X,Y\in\O{k}{N}
\;\;.\label{17}
\ee
The tangent space $T_{A}{\cal A}$ to $\cal A$ at a connection $A$ can be
identified with $\O{1}{N}$ and (\ref{17}) thus defines a metric
$g_{\cal A}$ on $\cal A$.
At each point $A\in\cal A$, $T_{A}{\cal A}$ can be
split into a vertical part $V_{A}=Im(d_{A})$ (tanget to the orbit of
$\cal G$ through $A$) and a horizontal part $H_{A}=Ker(d_{A}^{*})$ (the
orthogonal complement of $V_{A}$ with respect to the scalar product
(\ref{17})).
Explicitly this decomposition of $X\in\O{1}{N}$ into its vertical and
horizontal
parts is
\bea
X&=&d_{A}G_{A}^{0}d_{A}^{*}X + (X-d_{A}G_{A}^{0}d_{A}^{*}X)\;\;,\nonumber\\
 &\equiv& v_{A}X + h_{A}X\;\;,\label{18}
\eea
where $G_{A}^{0} = (d_{A}^{*}d_{A})^{-1}$ is the Greens function of the
scalar Laplacian (which exists if $A$ is irreducible). We will
identify the tangent space $T_{[A]}\C$ with $H_{A}$ for some representative
$A$ of the gauge equivalence class $[A]$.
Then $g_{\cal A}$ induces a metric $g_{\C}$ on $\C$ via
\be
g_{\C}([X],[Y])=g_{\cal A}(h_{A}X,h_{A}Y)\label{19} \;\;,
\ee
where $X,Y\in\O{1}{N}$ project to $[X],[Y]\in T_{[A]}\C$.
With the same notation the Riemannian curvature of $\C$ is
\bea
\langle\RC([X],[Y])[Z],[W]\rangle&=&
\langle *[h_{A}X,*h_{A}W],G_{A}^{0}*[h_{A}Y,*h_{A}Z]\rangle -
(X\leftrightarrow Y)\nonumber\\
&+ & 2 \langle *[h_{A}{W},*h_{A}{Z}],G_{A}^{0}*[h_{A}{X},*h_{A}{Y}]\rangle
\label{20}\;\;.
\eea
The last ingredient we would need to be able to write down the action
(\ref{2}) or (\ref{6}) is the Christoffel symbols of $g_{\C}$ or, rather,
particular components thereof. We will sketch the required calculation
below. Equipped with all this we can now exhibit the relation
between Donaldson theory and \sqm\ on $\A{3}$.

The action of Donaldson theory on a four-manifold $N$ in equivariant form
(i.e.~prior to the introduction of gauge ghosts) is \cite{ewdon}
\bea
S&=&\int_{N}\left(B_{+}F_{A}+\c_{+}d_{A}\p -B_{+}^{2}/2 +\e d_{A}*\p\right)
\nonumber\\
 & &+ \left(\fb d_{A}*d_{A}\f + \fb[\p,*\p] -\f [\c_{+},\c_{+}]/2\right)
\;\;.\label{21}
\eea
Here $F_{A}=dA+\frac{1}{2}[A,A]$ is the curvature of the connection $A$,
$\p\in\O{1}{N}$, the superpartner of $A$, is a Grassmann odd Lie algebra
valued one-form with ghost number 1, $(B_{+},\c_{+})$ are self-dual
two-forms with ghost numbers $(0,-1)$ (Grassmann parity (even,odd)),
and $(\f,\fb,\eta)$ are elements of $\O{0}{N}$ with ghost numbers $(2,-2,-1)$
and parity (even,even,odd). The equivariantly nilpotent BRST-symmetry of
(\ref{21}) is
\bea
\d A &=& \p\;\;\;\;\;\;\;\;\;\;\d \p = -d_{A}\f\nonumber\\
\d \c_{+}&=& B_{+}\;\;\;\;\;\d B_{+} = [\f,\c_{+}]\nonumber\\
\d \fb &=& \eta\;\;\;\;\;\;\;\;\;\;\;\d \eta = [\f,\fb]\nonumber\\
\d \f &=& 0 \;\;\;\;\;\;\;\;\;\;\;\d^{2}=\d_{\f}\label{22}
\eea
where $\d_{\f}$ denotes a gauge variation with respect to $\f$.
The action (\ref{21}) is far from being unique. In particular, by standard
arguments of \tft\, many $\d$-exact terms can be added to the action without
changing the partition function or
correlation functions (the Donaldson invariants in this case). We will make
use of this freedom below. For many of the other things that can and
should be said about (\ref{21},\ref{22}) we refer to \cite{ewdon} and
\cite[pp.~199-235]{pr}.

If $N$ is of the form $N=M\times S^{1}$ (where we think of $S^{1}$ as
the `time' direction) we can
perform a $(3+1)$-decomposition of the action. Identifying the self-dual
two-forms $B_{+}$ and $\c_{+}$ with (time-dependent) elements $B$ and $\pb$
of $\O{1}{M}$, reserving henceforth the notation $A$ for the spatial part
of the connection, and renaming $A_{0}\ra u$ and $\p_{0}\ra\eb$
we find that (\ref{21}) takes the form
\bea
S&=&\int_{M}\int dt\left( B*(\dot{A}-d_{A}u-*F_{A}) -B*B/2 +\pb d_{A}\p
+\dot{\p}*\pb + \fb d_{A}*d_{A}\f\right)\nonumber\\
&+&\left(u[\p,*\pb]+ \e d_{A}*\p + \eb d_{A}*\pb +\fb[\p,*\p] -\f[\pb,*\pb]/2
\right)\label{23}\;\;.
\eea
In going from (\ref{21}) to (\ref{23}) we have, for later convenience,
also subtracted the BRST exact term $(D_{0}\equiv \del_{0}+[u,\;\,])$
\[\d(\fb D_{0}\eb)=\e D_{0}\eb + \fb[\eb,\eb] -\fb D_{0}D_{0}\f\;\;.\]
We now perform the following elementary manipulations (Gaussian integrals):
\begin{itemize}
\item Integration over $\e$ and $\eb$ forces $\p$ and $\pb$ to be horizontal,
      $h_{A}\p=\p$, $h_{A}\pb=\pb$, i.e.~to represent tangent vectors to $\C$
\item Integraton over $\fb$ yields $\f=-G_{A}^{0}*[\p,*\p]$, giving rise to
      a term
      \[\frac{1}{2}\langle *[\pb,*\pb],G_{A}^{0}*[\p,*\p]\rangle\]
      in the action
\item The equation of motion for $u$ reads
      \[u=G_{A}^{0}(d_{A}^{*}\dot{A}+*[\p,*\pb])\]
      and plugging this back into (\ref{23}) one obtains
\[\frac{1}{2}\langle h_{A}(\dot{A}-*F_{A}),h_{A}(\dot{A}-*F_{A})\rangle
 +\frac{1}{2}\langle *[\p,*\pb],G_{A}^{0}*[\p,*\pb]\rangle +
  \langle *[\p,*\pb],G_{A}^{0}d_{A}^{*}\dot{A}\rangle\]
\end{itemize}
Putting all this together we see that the combination of Greens functions
appearing is precisely that entering the equation (\ref{20}) for the
curvature tensor $\RC$ while the kinetic term for the gauge fields is exactly
$g_{\C}([\dot{A}-*F_{A}],[\dot{A}-*F_{A}])/2$ (eq.~\ref{19}).
Recalling that $*F_{A}$ is the (automatically horizontal)
gradient vector field of the Chern-Simons functional $CS(A)$ and $*d_{A}$ its
second derivative, we find perfect
term by term agreement with the action $S_{X,V}=S_{\C,CS}$ (\ref{6})
provided that $\langle *[\p,*\pb],G_{A}^{0}d_{A}^{*}\dot{A}\rangle$
correctly reproduces the affine terms appearing in the second line of
(\ref{6}). That this is indeed the case can be seen by noting that the
variation of the metric $g_{\C}$ with respect to $A$ arises solely from the
variation of the projectors $h_{A}$. As $\p$ and $\pb$ are horizontal
the only variation that will therefore contribute is that of the first
$A$ in the vertical projector $d_{A}G_{A}^{0}d_{A}^{*}$ giving rise to
the above term when contracted with $\dot{A}$ as in
$\langle \pb,\nabla_{t}\p\rangle$. By the same argument
the affine term in $\nabla\del V$ does not contribute as
$*F_{A}$, $\p$ and $\pb$ are all horizontal and one of them will be
annihilated by a $d_{A}^{*}$ appearing in the variation of the metric.

In summary, we have seen thus far that the standard action (\ref{21})
of Donaldson
theory on a four-manifold of the form $M\times S^{1}$ is precisely the
quantum mechanics action (\ref{6}) on $\A{3}$ rewritten, as in \cite{aj},
in local form with the help of auxiliary fields. Conversely, the action
(\ref{21}), for which several other constructions are also available
\cite{pr}, could have been used to {\em derive} the
metric and curvature on $\C$ from those on ${\cal A}$. It is in this sense
that (\ref{23}), without the (model dependent) terms coming from the potential,
provides a realization $S_{{\cal A}\ra\C}$ of the Riemannian submersion action
$S_{Z\ra X}$ mentioned at the end of section 2. This part of the action is
universal, i.e.~common to all \sqm\ actions on $\C$, and is the counterpart
of the universal action of $N=2$ \tgt\ describing the Riemannian
geometry of $\C$ and discussed in \cite{btmq}.

The general strategy for
the construction of such actions, at least in the case of fibrations,
should now also be clear: one modifies the nilpotent BRST symmetry (\ref{susy})
to an equivariantly nilpotent symmetry squaring (as in (\ref{22})) to
translations along the
fibers parametrized by a new (ghost number 2) field $\f$. Rest as before.

Our next goal is to show how to obtain a \tgt\ of flat connections in $3d$
from the \sqm\ action $S_{\C,CS}$ in $3d$. This is straightforward since,
by the arguments of section 2, only the time-independent modes contribute to
the partition function of (\ref{23}) which is thus the same as the partition
function of the action\footnote{To arrive at this action, multiply (\ref{23})
by $1/\beta$ and scale $t$ by $\beta$ so that the circle has unit-radius.
Then $\beta$ will appear only in terms with time-derivatives. To eliminate
these, scale the non-constant modes of $A$ and $\p$ by $\beta$. In the limit
$\beta\ra 0$ these modes decouple and one is left with (\ref{24}).}
\bea
S_{\M}&=&\int_{M}\left( \frac{1}{2}F_{A}*F_{A}+\frac{1}{2}d_{A}u*d_{A}u
-d_{A}\fb*d_{A}\f +\pb d_{A}\p\right)\nonumber\\
&+&\left(u[\p,*\pb]+ \e d_{A}*\p + \eb d_{A}*\pb +\fb[\p,*\p] -\f[\pb,*\pb]/2
\right)\label{24}\;\;.
\eea
This is precisely the action obtained in \cite{btmq} as a field theoretic
realization of the Euler number $\c(\M)$ of the moduli space $\M=\M(M,G)$
of flat connections. From the present derivation of this action it is
obvious that $Z(S_{\M})=\c(\M)$ while in \cite{btmq} we verified this by
calculating the partition function.  To that end we integrate
over $\f$, $\fb$, $u$, $\e$, and $\eb$ as above to obtain $\RC$. To
evaluate the integral over the remaining fields $A$, $\p$, and
$\pb$ we expand them about their classical configurations.
%\bea
%A=A_{c}+A_{q}\;\;&,&\;\;F_{A_{c}}=d_{A_{c}}*A_{q}=0\;\;,\;\;
%d_{A_{c}}A_{q}\neq 0\;\;,\nonumber\\
%\psi=\psi_{c}+\psi_{q}\;\;&,&\;\;d_{A_{c}}\p_{c}=d_{A_{c}}*\p_{c}=0\;\;,
%\nonumber\\
%\pb=\pb_{c}+\pb_{q}\;\;&,&d_{A_{c}}\pb_{c}=d_{A_{c}}*\pb_{c}=0\;\;.
%\label{25}
%\eea
By standard arguments we may restrict ourselves to a one-loop approximation
and to this order the remaining terms in the action become
\be
\int_{M}\left( \frac{1}{2}F_{A}*F_{A}+\pb d_{A}\p\right)\ra
\int_{M}(\frac{1}{2}d_{A_{c}}A_{q}*d_{A_{c}}A_{q}+ [\pb_{c},\p_{c}]A_{q})\;\;.
\label{26}
\ee
where we can choose $F_{A_{c}}=0$ as the coefficient of $B^{2}$ in (\ref{23})
is arbitrary. Integration over $A_{q}$ yields
\be
\frac{1}{2}\langle *[\pb_{c},\p_{c}],G_{A_{c}}^{1}*[\pb_{c},\p_{c}]
\rangle\;\;,\label{K2}
\ee
where $G_{A_{c}}^{1}$ is the Greens function of the Laplacian
$d_{A_{c}}d_{A_{c}}^{*}+d_{A_{c}}^{*}d_{A_{c}}$ on one-forms,
composed with a projector
onto the orthogonal complement of the space of $d_{A_{c}}$-harmonic
one-forms. Now the extrinsic curvature of $\M\ss\A{3}$ is
\[K_{\M}([X],[Y])=-d_{A}^{*}G_{A}^{1}*[\bar{X},\bar{Y}]\]
(see \cite{gp,btmq}) with $\bar{X}$ and $\bar{Y}$ satisfying the
linearized flatness and horizontality equations
$d_{A}\bar{X}=d_{A}^{*}\bar{X}=d_{A}\bar{Y}=d_{A}^{*}\bar{Y}=0$.
Therefore the above term (\ref{K2})
is precisely the $K_{\M}^{\;2}$ contribution to the Gauss equation
(\ref{10}) for $\RM$ and the partition function of $S_{\M}$ is indeed
the Euler characteristic $\c(\M)$.

We want to draw attention to the double-role played by the multiplier
field $B$: its exact part couples to $u$ and gives rise
to the submersion action $S_{{\cal A}\ra\C}$ and the O'Neill equations,
while its coexact part couples to the gradient $*F_{A}$ of the potential
and is responsible for the $K$-part of the Gauss-Codazzi equations.

The reader may be puzzled at this point by the fact that Donaldson theory
on $M\times S^{1}$ apparently calculates the Euler number of the moduli space
$\M(M,G)$ of flat connections in three dimensions although it is known
(and was invented) to describe moduli spaces of instantons in four dimensions.
How this happens is explained in detail in \cite{btmq}. It is essentially
due to the fact that the index of the instanton deformation
complex (the formal dimension of the instanton moduli space) on a four-manifold
of the form $M\times S^{1}$ is non-zero in the topologically non-trivial
sector so that the partition function vanishes there, while irreducible
`instantons' in the trivial sector correspond to flat connections in $3d$.

We end this section with the remark that,
by a result of Taubes \cite{tau}, the partition function
of (\ref{24}) formally equals the {\em Casson invariant} of $M$ if $M$ is
a homology three-sphere \cite{ewtop}. This, combined with the above
considerations, has led us to propose $\c(\M)$ as a candidate for the
definition of the Casson invariant of more general three-manifolds
(see \cite{btmq} for some preliminary considerations).

\section{Other examples}

We will now briefly discuss the corresponding constructions for those
moduli spaces $\M$ which are not of the form $X_{V}$
for some potential $V$ on $X=\C$.
In that case we will, as discussed in section 2, use the
Gauss-Codazzi \sqm\ action $S_{\M\ss\C}$ (\ref{6}) which will be the sum of
$S_{\C}$ (or its local counterpart $S_{{\cal A}\ra\C}$), and
the $N=2$ supersymmetric delta function constraints onto $\M\ss\C$.
Alternatively, to construct e.g.~the action associated to the moduli
space $\M_{2}$ of flat connections in two dimensions we can simply
dimensionally reduce the action (\ref{23}) assuming that
$M=\Sigma\times S^{1}$. This makes the double role
played by $B$ particularly transparent: the action $S_{\A{3},CS}$ reduces
to the Gauss-Codazzi action $S_{\M\ss\A{2}}$ via
\bea
S_{\A{3},CS}&=&\int B*(\dot{A}-d_{A}u-*F_{A}) + \ldots \nonumber\\
\ra S_{\M_{2}\ss\A{2}}&=& \int B*(\dot{A}-d_{A}u) +b F_{A} +\ldots
\label{27}
\eea
($b$ is the scalar (time) component of $B$). Now $B$ evidently
represents the submersion (O'Neill) part of the action, while $b$
represents the embedding (Gauss-Codazzi) part. Proceding as above
to extract the two-dimensional (zero mode) action from (\ref{27}) one
finds
\bea
S_{\M_{2}}&=&\int_{\Sigma}\left(bF_{A} +\frac{1}{2}d_{A}u*d_{A}u+u[\p,*\pb]
-d_{A}\fb*d_{A}\f +\cb d_{A}\p-\c d_{A}\pb + \frac{1}{2}B*B\right)\nonumber\\
&+&\left(\e d_{A}*\p + \eb d_{A}*\pb +\fb[\p,*\p] -\f[\pb,*\pb]/2
+u(d_{A}B+[\p,\pb])\right)\;\;.\label{28}
\eea
Once again, this is precisely the action derived in \cite{btmq} satisfying
$Z(S_{\M_{2}})=\c(\M_{2})$. This can, of course, also be established by
direct calculation. This calculation is somewhat simpler here than in three
dimensions as we have a delta function constraint instead of a Gaussian
around $\M_{2}$ so that the partition function can be calculated directly
without performing a classical-quantum split. In particular,
it is now the integration over $u$ and $B$ that will produce the extrinsic
curvature contribution which is (cf.~(\ref{K2}))
$\frac{1}{2}\langle *[\pb,\p],G_{A}^{0}*[\pb,\p]\rangle$, in agreement
with the calculations in \cite{gp,btmq}.

We also want to mention that the non-degeneracy
condition $det(F^{ab})\neq 0$ we encountered in our discussion of
Gauss-Codazzi \sqm\ in section 2 is just the condition that the Laplacian
on zero-forms be invertible, i.e.~that the connection be irreducible,
as we have assumed all along. It is, of course, only at those points that
the condition $F_{A}=0$ gives a non-singular description of $\M_{2}$.
It appears likely that a non-singular description of the reducible points
(automatically gauge fixing the residual symmetry there) can be obtained by
adding a term $\gamma\Lambda^{2}$ (\ref{12}) to the action
$S_{\M\ss\C}$ (\ref{13}) and carefully taking the limit $\gamma\ra 0$.
This, as well as the other suggestions for dealing with reducible connections
put forward in \cite{btmq}, is currently under investigation.

A related issue is the question, what the path integral calculates if the
target spaces $X$, $Y\ss X$ or $X_{V}\ss X$ (and, in particular, the
moduli spaces $\M\ss\C$) are not smooth manifolds but perhaps orbifolds
or orbifold stratifications. In the case of orbifolds one expects to
obtain the vitual Euler characteristic via Satake's Gauss-Bonnet
theorem for $V$-manifolds \cite{satake}. On the other hand, the equivariant
orbifold Euler characteristic familiar from string theory
appears to arise upon reduction of \sqm\ on $X=Y\times S^{1}$ to $Y$
with twisted boundary conditions on both the `temporal' and `spatial'
circles.

Finally, in order to obtain a theory modelled on the moduli spaces
$\M_{I}$ of instantons in four dimensions, we can construct the
corresponding Gauss-Codazzi quantum mechanics action on $\A{4}$. The
construction is almost identical to that for $\M_{2}$ (essentially
because the deformation complex is `short' in both examples so that no
additional gauge fixing is required) , and the resulting $4d$ action
is obtained from (\ref{28}) by replacing the scalar $b$-multiplet by a
multiplet of self-dual two-forms.
\begin{center}
{\bf Acknowledgements}
\end{center}
We thank the Bundesministerium f\"ur Forschung und Technologie (Bonn,FRG)
and the Stichting voor Fundamenteel Onderzoek der Materie (Utrecht, NL)
for financial support.

\end{document}